\documentclass[nature,aps,unsortedaddress,amsmath,amssymb]{revtex4}
\usepackage{graphicx}
\usepackage{amsmath}
\usepackage{amssymb}
\usepackage{color}

\begin{document}
\title{Acoustic-like dynamics of amorphous drugs in the THz regime.}
\author{%
E.A.A. Pogna$^{1}$,
C. Rodr\'iguez-Tinoco$^{2}$,
M. Krisch $^{3}$,
J. Rodr\'iguez-Viejo $^{2,4}$,
T. Scopigno$^{1}$}
\affiliation{$^{1}$Dipartimento di
Fisica,~Universit\'a~di~Roma~"La Sapienza",~I-00185,~Roma,~Italy
\email{tullio.scopigno@phys.uniroma1.it}
}
\affiliation{$^{2}$Nanomaterials and Microsystems Group, Physics Department,
Universitat Aut$\grave{o}$noma de Barcelona, ES 08193,~Bellaterra,~Spain
}
\affiliation{$^{3}$European Synchrotron Radiation Facility, Boîte Postal 220, F-38043 Grenoble, France
}
\affiliation{$^{4}$MATGAS 2000 AIE, Campus UAB, 08193 Bellaterra, Spain
}
\date{\today}
\begin{abstract}

The high frequency dynamics of Indomethacin and Celecoxib glasses has been investigated by inelastic x-ray scattering, accessing a momentum-energy region still unexplored in amorphous pharmaceuticals. We find evidence of phonon-like acoustic dynamics, and determine the THz behavior of sound velocity and acoustic attenuation. Connections with ordinary sound propagation are discussed, along with the relation between fast and slow degrees of freedom as represented by non-ergodicity factor and kinetic fragility, respectively.
\end{abstract}

\maketitle
Recently, the preparation of pharmaceutical substances in glassy state has been proven successful
in improving their therapeutic performances \cite{murdande_solubility_2011,hancock_characteristics_1997}.
The glassy state offers advantages
such as higher solubility, dissolution rate, and
sometimes better compression properties as compared
to the crystalline form. As models of amorphous drugs, Indomethacin and Celecoxib are of particular interest because in
their crystalline form they are both limited by poor water-solubility
\cite{swallen_self-diffusion_2011,gupta_physical_2004,grzybowska_molecular_2010-1,chawla_characterization_2003,paradkar_preparation_2003}.
At the microscopic scale, the glassy state is characterized by the lack of long range structural order and it exhibits thermodynamical and transport properties consistently
different from those of the crystalline phase.
Extensive research in the past decades has demonstrated that all glasses,
without sensitivity to their composition, exhibit identical thermodynamic anomalies,
the most notorious being the thermal conductivity plateau and specific heat C(T) excess over the Debye expectation in the low-temperature regime, near 10K \cite{zeller_thermal_1971-1}.
These universal features are generally ascribed to an excess in the density of vibrational states over the trend $\propto\omega^2$ predicted by the Debye model for crystalline solids \cite{binder_glassy_2005}, conventionally called Boson peak (BP).
The Boson peak is experimentally observed in all glasses by Raman \cite{schulte_inelastic_2011,schmid_raman_2008} and inelastic neutron scattering \cite{bove_brillouin_2005}, and numerically reported by molecular dynamics simulations and in focus of several theoretical efforts \cite{marruzzo_vibrational_2013,singh_ultrastable_2013,derlet_boson_2012,marruzzo_heterogeneous_2013,_conference_2012}.
The anomalies of the density of state can be, in turn, inferred from the peculiarities of the frequency spectra of atomic vibrations.
While the vibrational properties of the crystalline state are well rationalized in terms of phonon modes, those of the glassy state are still a matter of intense debate.
Due to the lack of translational
invariance, even in "harmonic" glasses, the vibrational eigenstates are distinctly different from ideal plane waves.
Nevertheless, since they exhibit propagating behavior in the long wavelength limit, they are conventionally described by the phonon-like formalism
inherited from crystals.
The development of high resolution (meV) inelastic x-ray scattering (IXS) \cite{sette_dynamics_1998-2} has opened the possibility of investigating acoustic dynamics of glasses in the BP region. A sizable library of IXS observations indicates the existence of vibrational excitations characterized by well defined dispersion curves resembling those of the crystalline counterparts, though with distinctive properties influenced by disorder, such as the complex sound attenuation dependencies on temperature and frequency. Acoustic attenuation in glasses is indeed determined by the interplay of
different mechanisms \cite{rayleigh_theory_1877,maris_6_1971,ferrante_acoustic_2013}. In the high (THz) frequency regime, i.e. at acoustic wavelengths comparable to the mean interparticle distances, the temperature independent effect of structural disorder dominates and it is responsible for a progressive localization of acoustic modes occurring as consequence of scattering.
One criterion to determine the localization threshold is the
Ioffe-Regel criterion, i.e. the equivalence between the excitation mean free path and the dominant excitation wavelength. It has been proposed that the Ioffe-Regel condition on longitudinal vibrational modes coincides with the BP position \cite{ruffle_glass-specific_2006}, although this is not generally accepted \cite{ruocco_comment_2007} and it has been pointed out that the criterion should be rather applied to the transverse acoustic branch \cite{marruzzo_heterogeneous_2013,shintani_universal_2008,schober_vibrations_2004,monaco_anomalous_2009}.
In recent years, the idea that vibrational properties of glasses may be related to the liquid side of the glass formation process has gradually emerged.
Specifically, a strong correlation between the temperature dependence of the non-ergodicity factor, and the kinetic fragility m of the corresponding liquids
has been proposed as universal feature of glass formation \cite{scopigno_is_2003,scopigno_universal_2010,niss_glassy_2008}. The former represents the long time limit of the normalized density-density correlation function \textit{in the glass}, it is therefore a measure of the decorrelation introduced by the fast vibrational dynamics or, equivalently, of the residual correlation over the ergodicity of the underlying liquid phase. The latter is the steepness of the (supercooled) liquid viscosity measured at the glass transition, i.e. as the diffusive motion of the liquid state slows down \cite{bohmer_nonexponential_1993,angell_structural_1988}. Liquids whose fragility m ranges from 17 to 30 are classified as strong liquid (Arrhenius-like behavior of viscosity), such as silica, whereas fragility values from 100 to 150 correspond to so-called fragile systems (T-dependent and non-Arrhenius like behavior of viscosity), such as o-terphenyl.
Remarkably, the proposed correlation of kinetic fragility to the non-ergodicity factor in the low temperature T = 0 limit can be used to introduce a definition of fragility uniquely from glass properties away from T$_{g}$.
Fragility correlates, in turn, to several other essential properties in the glass, such as molecular mobility below T$_{g}$ that is believed to be an important parameter to explain the physical stability of some amorphous compounds towards crystallization \cite{yoshioka_correlations_2007,baird_role_2012}. In fragile liquids
 the molecular mobility changes rapidly approaching T$_{g}$ and this is why 'strong' glass-formers are considered to be more stable than 'fragile' ones \cite{angell_structural_1988}, a property well known to glass blowers since a long time.
Understanding the critical factors governing the crystallization tendency of organic compounds is of fundamental interest when assessing the feasibility of an amorphous formulation for pharmaceuticals. The dependence on the method of preparation has been consistently explored, for example, determining the crucial role played by the cooling rate in glasses synthesized by quench cooling of the liquid \cite{Karmwar_investigation}.
Many amorphous drugs, IMC and CXB included, are characterized by large
values of kinetic fragility that classify them as fragile materials and they show a large tendency towards crystallization that prevents their continuous utilization in the amorphous form \cite{yu_amorphous_2001-1,yoshioka_crystallization_1994}.
High frequency vibrational properties of amorphous drugs are still unexplored. We report here IXS measurements on two pharmaceutical glasses with different kinetic fragility, Indomethacin and Celecoxib.
Specifically, we determine the longitudinal sound dispersion, the high frequency acoustic attenuation and the non-ergodicity factor, discussing the eigenmodes localization and their connection with mass diffusivity.
\section*{Results}
\begin{figure}[!htbp]
\centering
\includegraphics[width=0.5\textwidth]{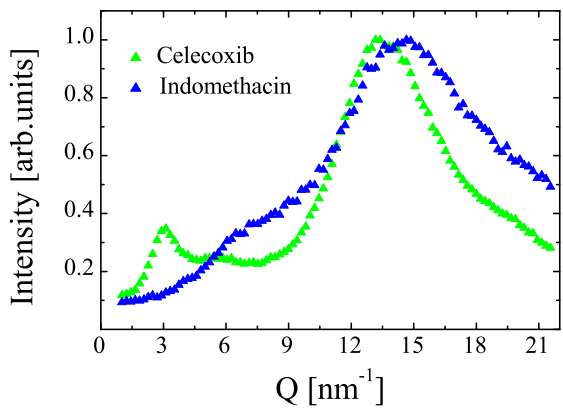}
\caption{X-ray diffraction patterns of Indomethacin (blue triangles) and Celecoxib glasses (green triangles) as function of the exchanged wave vector Q, measured at room temperature $300K$ using an incident energy $E=23.725~KeV$, normalized to the intensity of the main peak.}
\label{f:static}
\end{figure}

\begin{figure}[!htbp]
\includegraphics[width=0.45\textwidth]{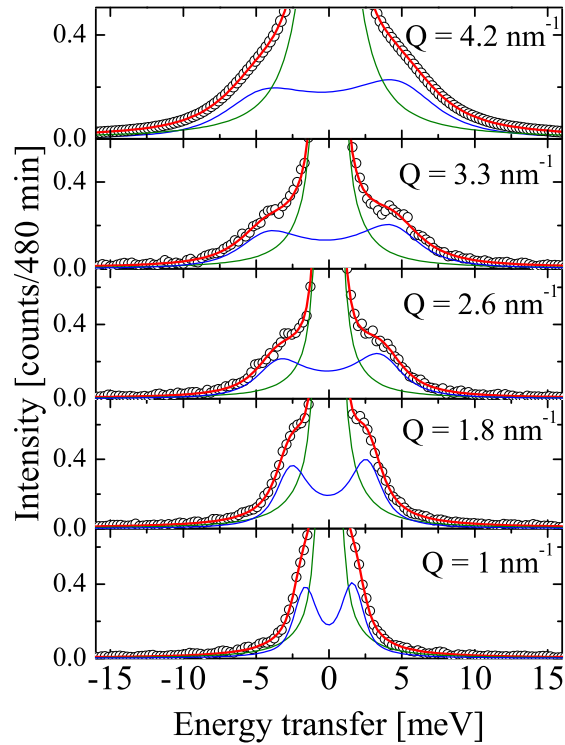}
\caption{Selection of room temperature inelastic x-ray spectra of glassy Indomethacin at fixed exchanged momentum Q along with the best fits (red line) described in section Methods. The energy resolution (green line) normalized to the elastic peak (elastic component) is also reported, while the blue line corresponds to the inelastic, damped harmonic oscillator contribution. The experimental error is within the symbol.}
\label{imc_spectra}
\end{figure}

\begin{figure}[!htbp]
\centering
\includegraphics[width=0.5\textwidth]{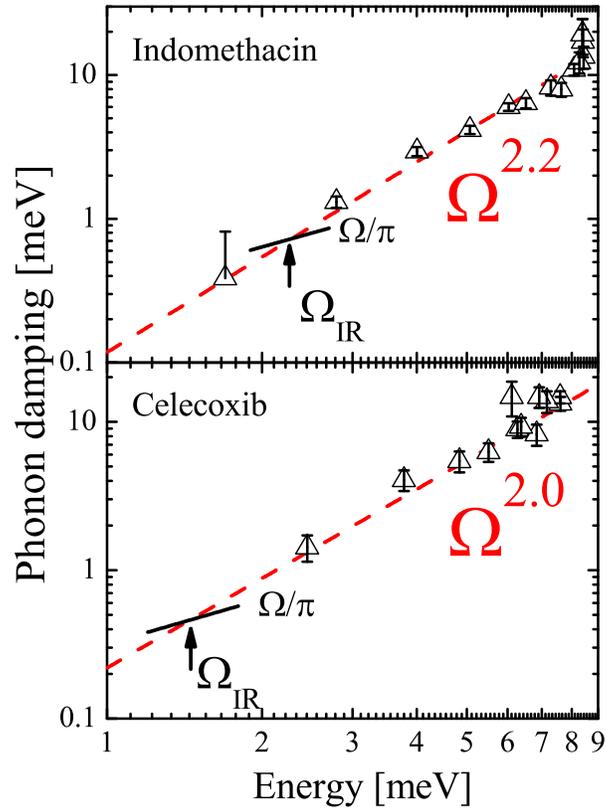}
\caption{Acoustic attenuation $\Gamma$ of Indomethacin (upper panel) and Celecoxib (lower panel) as function of energy of the acoustic mode $\Omega$ on a log log scale, along with the best fit result utilizing a power law of exponent 2.2 for IMC and 2.0 for CXB. The Ioffe-Regel condition $\Omega_{IR}$ is determined from the crossing frequency of the attenuation curves with the function $\Omega/\pi$.}
\label{f_attenuation}
\end{figure}

The x-ray diffraction measurements on Indomethacin (IMC) and Celecoxib (CXB),
 detected with a photodiode embedded in the inelastic analyzers bench, show a broad main peak centered at Q$_{p}^{CXB}$=13.2nm$^{-1}$ and Q$_{p}^{IMC}$=14.7nm$^{-1}$,
confirming the amorphous nature of the samples (see Fig.\ref{f:static}).
A selection of inelastic x-ray scattering spectra of IMC in the glassy state are shown in Fig.\ref{imc_spectra} at five fixed values of exchanged momentum Q. They are dominated by an intense elastic feature due to the non-ergodicity of the glass, i.e. to density fluctuations frozen by the structural arrest.
At both sides of the elastic component, well defined inelastic peaks
are observed at finite energy transfer.
The energy position $\Omega$ of the inelastic peaks corresponds to the energy of the acoustic modes of momentum Q, while the peak width gives access to the life time of density fluctuations , i.e. to the sound attenuation $\Gamma(Q)$.
The IXS spectrum can be modeled by a delta function, describing the elastic component, combined with the power spectrum of a damped harmonic oscillator (DHO, corrected to account for the detailed balance condition) representing the inelastic scattering.
Both contributions have been convoluted with the instrument response function (green line in Fig.\ref{imc_spectra}, for further details see the Methods section) to correctly reproduce the experimental data.
We note that the inelastic contributions to the total scattered intensity due to the Brillouin inelastic scattering from vibrational density fluctuations
are well reproduced (blue line in Fig.\ref{imc_spectra}).
The energy position of the Brillouin component increases with exchanged momentum Q, testifying the propagating nature of the probed excitations. This general finding proves the existence of THz hypersonic waves in amorphous IMC and CXB. Since only one inelastic contribution is resolved at each Q value, the vibrational excitations can unambiguously be ascribed to a single longitudinal acoustic phonon branch.
The inelastic component becomes less and less visible as the exchanged momentum increases, due to an increase of the acoustic attenuation (broadening of the inelastic features) and to the increase of the elastic component which follows the static structure factor S(Q) (see Fig.\ref{f:static}).
\begin{figure}[!htbp]
\centering
\includegraphics[width=0.5\textwidth]{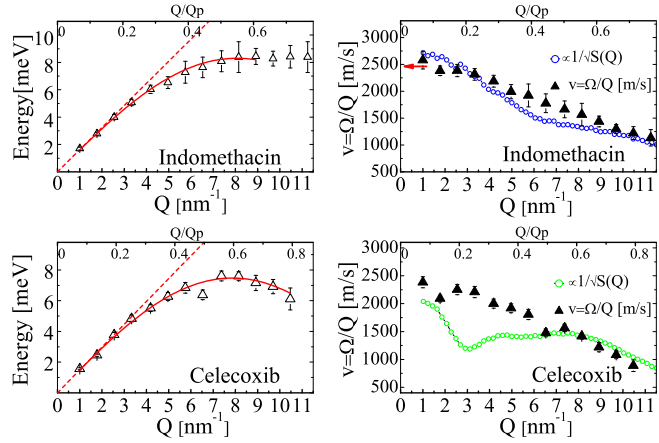}
\caption{Energy peak position $\Omega(Q)$, as obtained from the fit of the data to eq.\ref{fitS}, for (a) Indomethacin
(b) Celecoxib measured as function of exchanged momentum Q. The parameter $\Omega(Q)$ corresponds
to the maximum of the longitudinal current that is $\omega^2/Q^2$S(Q,$\omega$). The bottom axes represent the exchanged momentum Q; the top axes show Q/$Q_{p}$ where $Q_{p}$ is the maximum of the XRD patterns. Apparent sound velocity $v=\Omega/Q$ along with its expected hydrodynamic Q-dependency $\propto 1/\sqrt{S(Q)}$ in (c) Indomethacin (d) Celecoxib.}
\label{f_dispersion}
\end{figure}
The energy dispersion curves, $\Omega(Q)$, of the longitudinal acoustic phonon are linear in the long wavelength-limit Q $\leq$ 3nm$^{-1}$; they then bend down reaching a maximum at Q around half of Q$_{p}$, highly suggestive of the existence of a pseudo-Brillouin zone (see Fig.\ref{f_dispersion}). The plateau in the phonon dispersion of Indomethacin indicates the end of the propagating behavior while in the Celecoxib the group velocity remains finite up to 0.8 Q$_{p}$.
The acoustic attenuation  $\Gamma$
shows for both samples a power law dependence on the acoustic frequency $\Omega(Q)$ compatible with
the quadratic law found in different classes of glassy systems in a similar frequency regime (see Fig.\ref{f_attenuation}).
\section*{Discussion}
\begin{figure}[!htbp]
\centering
\includegraphics[width=0.5\textwidth]{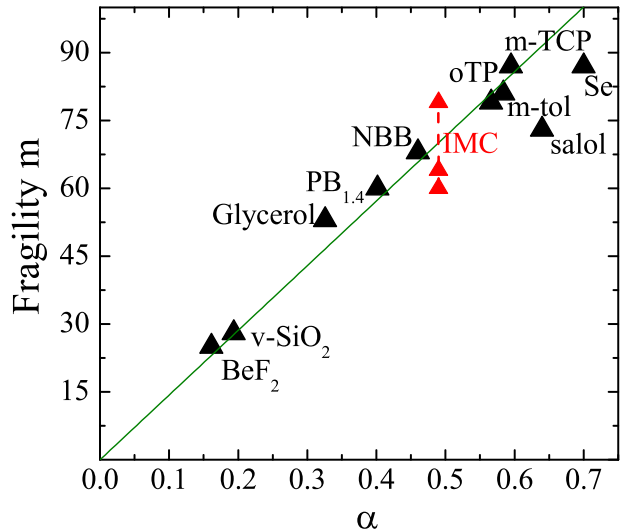}
\caption{Correlation plot of the kinetic fragility m and the $\alpha$ parameter of the non-ergodicity factor for several glassy systems. Three different experimental estimations of fragility available in literature are reported for IMC, ranging from 60 to 79 \cite{ramos_activation_2002,correia_molecular_2001,swallen_self-diffusion_2011}.}
\label{f_fragility}
\end{figure}

The interest in amorphous form of Indomethacin and Celecoxib is relatively recent and there is no report of previous characterization
of the hypersonic sound propagation.
The energy dispersion curves reported here have been analyzed with a sinusoidal fitting function (solid red line in Fig.\ref{f_dispersion}).
In the case of Indomethacin we used Q values up to Q/Q$_{p}=0.6$
because for higher exchanged momentum there is no longer any significant dispersion.
From the linear fit of the dispersion curves in the limit Q$\rightarrow 0$ (dashed red line in Fig.\ref{f_dispersion}) the longitudinal sound velocity in Indomethacin and Celecoxib has been estimated to be equal to $2460~m/s$ and $2214~m/s$, respectively.
In the case of IMC glass, previous Brillouin Light Scattering measurement at few GHz exists \cite{kearns_high-modulus_2010} (red arrow in the sketch of upper panel in Fig.\ref{f_dispersion}). The long wavelength extrapolation of the hypersonic velocity reported here smoothly approaches the light scattering results indicating that, within our experimental accuracy, we do not detect anomalous (positive or negative) dispersion effects reported in different glasses \cite{monaco_breakdown_2009-1,monaco_anomalous_2009,ruta_acoustic,ruzicka_positive,scopigno_microscopic_2005}. A similar comparison for CXB glass is not possible due to the lack of any sound velocity data.
The sound velocity dispersion $v_{a}(Q)=\frac{\Omega(Q)}{Q}$ was compared with the hydrodynamic prediction $1/\sqrt{S(Q)}$ expected by a Langevin equation scenario according to which, for a monoatomic system of particles of mass m, $v_{a}(Q)$ reads:
\begin{equation}
v_{a} = \sqrt{\frac{K_{B}T}{ mS(Q)}}
\end{equation}
The role of a single particle in a monoatomic system, is here played by a molecular unit. The result is reported in Fig.\ref{f_dispersion} over the accessed Q region.
Indomethacin verifies quite well the prediction, while in CXB there is a discrepancy in the proximity of the pre-peak at Q$\approx 3$nm$^{-1}$, indicating that this latter structural feature is not connected to collective dynamics.
As we pointed out, microscopic disorder induces the localization of vibrational excitations signaled by the Ioffe-Regel crossover.
The onset frequency $\Omega_{IR}$ is defined as the inverse of
the decay time of the plane wave $\tau=\frac{1}{\pi\Gamma}$ such that:
\begin{equation}
\Gamma_{IR}=\frac{\Omega_{IR}}{\pi}
\end{equation}
The Ioffe-Regel crossover frequencies determined by longitudinal vibrational excitations of IMC and CXB are found at
$\Omega_{IR}^{IMC}= 2.3~meV$ and $\Omega_{IR}^{CXB}= 1.4~meV$, respectively (see Fig.\ref{f_attenuation}).
These values are compatible with that of other molecular glasses \cite{ruffle_glass-specific_2006}. Remarkably, in the case of IMC and CXB the Ioffe-Regel crossover determined for longitudinal waves does not correspond to spatial localization of vibrational excitations since we observe the existence of low energy dispersive excitations up
to almost $\Omega=~8~meV$ ($Q=~7~nm^{-1}$). According to simulations on vitreous silica \cite{taraskin_low-frequency_1999},
the Ioffe-Regel crossover frequency for vibrational excitations may not correspond to mode
localization but rather to the change into diffusive,
spatially extended states, resulting
from the mixing of propagating phonons
with localized vibrational modes.
As discussed in the introduction, the sound attenuation in glasses displays a complex behavior when studied over an extended wavelength range. In this study, a nearly quadratic dependence $\Gamma$ = $a~\nu^{2.2}$ in the THz domain is reported for IMC with a coefficient $a=0.7~THz^{-1}$. Quadratic dependencies with different coefficients for the low frequency, anharmonic, behavior and the high frequency, disorder dominated, regime have already been reported in literature in strong glass-formers \cite{ferrante_acoustic_2013,ruffle_glass-specific_2005}, with an higher coefficient for the latter. Interestingly, from a previous Brillouin scattering study in IMC in the GHz range at 325 K one would get a coefficient $13.5~THz^{-1}$ \cite{kearns_high-modulus_2010}, i.e. the opposite trend is observed, suggesting a larger contribution from anharmonicity.
In order to pursue the correlation between mechanical properties and thermodynamical features
of the two investigated drugs, the non-ergodicity factor
$f(Q,T)$ is estimated, according to eq.\ref{nef}, as the ratio between
the elastic and total intensity of IXS spectra :
\begin{equation}
f(Q,T)=\frac{I_{el}}{I_{el}+I_{inel}}
\label{nef}
\end{equation}
Invoking the harmonic approximation
for the vibrational dynamics, the temperature dependence
of the non-ergodicity factor in the low Q region can be described \cite{scopigno_is_2003} as:
\begin{equation}
f(Q\rightarrow 0,T)=\frac{1}{1+\alpha\frac{T}{T_{g}}}
\label{approx}
\end{equation}
where the coefficient $\alpha$ contains all the microscopic details
of the system.
From equations \ref{nef} and \ref{approx} the coefficient $\alpha$ can be written as:
\begin{equation}
\alpha=\frac{I_{inel}}{I_{el}}\frac{T_{g}}{T}
\label{alfa}
\end{equation}
The estimated parameter $\alpha$ of IMC was compared to that of several glassy systems and the proposed correlation with kinetic fragility verified (see
Fig.\ref{f_fragility}).
IMC is usually classified as fragile material, though determination of m by different methods (dielectric spectroscopy, calorimetry and diffusivity) are scattered in the range $60<m<79$ \cite{ramos_activation_2002,correia_molecular_2001,swallen_self-diffusion_2011}. Based on the correlation with the non-ergodicity factor the extrapolated fragility of IMC is $m_{\alpha}=70$, a value that falls within the range of previous estimations, very close to the calorimetry measurement.
A reliable estimate of the non-ergodicity factor in Celecoxib could not be obtained due to the pre-peak in the static form factor (see Fig.\ref{f:static}). The latter has no dynamical origin, as it is shown by the disagreement between the generalized sound velocity and the 1/$\sqrt(S(Q))$ behavior (see Fig. \ref{imc_spectra}) and it is responsible for an additional contribution to the quasi elastic scattering preventing the determination of the non ergodicity factor with the aforementioned method. Summing up, this study provides a complete characterization of the THz dynamics of two prototypical pharmaceuticals prepared in the glassy state. Our results demonstrate the existence of acoustic-like, hypersonic vibrational excitations with a well defined dispersion relation and disorder dominated damping. The measure of the non-ergodicity factor, as determined by IXS spectra, is utilized for assessing the fragility of the glass phase. On the basis of Ref. \cite{yoshioka_correlations_2007}, we anticipate that the latter, in turn, may influence the molecular mobility of the supercooled phase, a parameter that has been proposed to rule the glass-forming ability of a given compound, and thus that has fundamental importance for amorphous pharmaceuticals."
\section*{Methods}
Indomethacin (99$\%$ purity, $T_{g}$= $315$ K and $T_{m}$($\gamma$ form)= $428$ K) and Celecoxib (98$\%$ purity with $T_{g}$= $326$ and $T_{m}$= $435$ K) crystalline powders were purchased from Sigma-Aldrich and BOC-Science, respectively.
Samples powders were placed onto a silicon substrate, shaped in order to guarantee the optimal x-ray path length $L$, such that it matches the photoelectric absorption length, $L=\mu^{-1}$, where $\mu$ is the absorption coefficient.
The silicon substrate was beforehand covered with a 20 nm aluminum layer to improve adhesion.
The powders were slowly heated above their melting temperature $T_{m}$, while slowly spinning the substrate to improve thickness homogeneity.
Once the sample was an homogeneous and well transparent liquid, the substrate was removed from the heater and self-cooled at $100~K/min$ to room temperature. During cooling the liquid freezes into a disordered glassy state at $T_{g}$.
The thickness of the resulting glasses were several mm. In order to prevent potential moisture absorption, the samples were stored in vacuum sealed boxes and the experiment was performed under dynamic high vacuum condition ($10^{-6}$bar). Control experiments, in inert gas and ambient conditions using differential scanning calorimetry, confirm moisture absorption during processing was negligible.
The IXS experiment was carried out at the beam line ID28 of the European Synchrotron Radiation Facility (ESRF). Measurements of the scattered intensity
at fixed scattering angles,
and therefore fixed exchanged momentum Q values, were performed at room temperature
($T=~300K$). An eight-analyzer bench, operating in horizontal
scattering geometry, allowed the simultaneous collection
of spectra at eight different values of Q. For both samples, the explored momentum-region
 extended from $1$ up to $14.6~nm^{-1}$
thus reaching the position, Q$_{p}$, of the main peak in
the diffraction pattern
of both samples.
The Q resolution determined by slits placed in front of the analyzer was set to $0.25~nm^{-1}$.
The scanned energy range was
$-23~\leq\Omega\leq~23~meV$, where $\Omega$ is the energy transfer such that
$\Omega=E_0 - E$, with $E_0$ and $E$ being the energy of the
incident ($23.725~eV$) and the scattered x-ray photon; each scan took approximately
$480~min$.
Using the (12 12 12) reflection for the Si monochromator
and crystal analyzers
the overall energy resolution was $1.4~meV$, the FWFM of the green line in Fig.\ref{imc_spectra} superimposed to the elastic peak.
The fitting function adopted to analyze the IXS spectra is the result of
the convolution of the model function:
\begin{equation}
\begin{split}
&S(Q,\Omega)=S(Q)[ 2\pi\delta(\Omega)f_{Q}]+ \\ &(1-f_{Q})\left[\frac{2\Omega(Q)\Gamma(Q)}{[\Omega^2-\Omega(Q)^2]^2+\Omega^2\Gamma^2(Q)} \right]\frac{\hbar\Omega/(k_{B}T)}{1-e^{\hbar\Omega/(k_{B}T)}}
\end{split}
\label{fitS}
\end{equation}
with the instrument resolution.
The two terms in eq.\ref{fitS} represent respectively the elastic
and inelastic part of the spectra, where $\Omega$ defines the position of the inelastic peaks and $\Gamma$ their widths.
The assumption of a delta function for the elastic component is justified because the experimental data in the quasi-elastic region do not show any broadening compared to the instrument resolution. The inelastic component is described by a damped harmonic oscillator that is the solution of the generalized hydrodynamics
using a Markovian memory function, i.e. assuming instantaneous relaxation of the molecular
vibrations. This has been shown to be appropriate for the glassy state \cite{scopigno_microscopic_2005}.
In order to account for the quantum nature
of the probed excitations, we multiplied the function by the Bose factor
such that it is adapted to satisfy the detailed balance condition.
\bibliographystyle{nature}

\section*{Acknowledgments}
E.A.A.P. and T.S. have received funding from the European Research Council under the European Community's Seventh Framework Program (FP7/2007–2013)/ERC grant Agreement No. 207916. C.R.T and J.R.V. acknowledge financial support from Generalitat de Catalunya and Ministerio de Economía y Competitividad through grants SGR2009-01225 and MAT2010-15225, respectively.
\section*{Author contributions}
T.S. conceived and supervised the research. E.A.A.P. C.R.-T. M.K. and T.S performed the experiment. E.A.A.P. analysed the data. C.R.-T. and J.R.-V. provided and characterized the samples. E.A.A.P. and T.S. wrote the manuscript. All the authors participated to the discussion of the results and commented on the manuscript.
\section*{Additional information}
\textbf{Competing financial interests:} The authors declare no competing financial interests.
\end{document}